\documentclass[aps,prc,amsfonts,superscriptaddress,showpacs,twocolumn,byrevtex,nofootinbib]{revtex4}

\usepackage{epsfig}
\usepackage{amsmath}
\usepackage{amssymb}
\usepackage{amsfonts}
\usepackage{color}

\allowdisplaybreaks     

\newcommand{\JMcomm}[1]{{\textcolor{black}{ #1}}}
\newcommand{\MBcomm}[1]{{\textcolor{black}{ #1}}}

\newcommand{\old}[1]{{\textcolor{black}{ }}}

\renewcommand{\vec}[1]{\mathbf{#1}}

\begin{document}

\title{Density functional theory and Kohn-Sham scheme for  
       self-bound systems}

\author{J{\'e}r{\'e}mie Messud}
\affiliation{
Universit\'e de Toulouse; UPS; Laboratoire de Physique
  Th\'eorique (IRSAMC); F-31062 Toulouse, France
}
\affiliation{
CNRS; LPT (IRSAMC); F-31062 Toulouse, France
}

\author{Michael Bender}
\affiliation{Universit{\'e} Bordeaux,
             Centre d'Etudes Nucl{\'e}aires de Bordeaux Gradignan, UMR5797,
             F-33175 Gradignan, France}
\affiliation{CNRS/IN2P3,
             Centre d'Etudes Nucl{\'e}aires de Bordeaux Gradignan, UMR5797,
             F-33175 Gradignan, France}

\author{Eric Suraud}
\affiliation{
Universit\'e de Toulouse; UPS; Laboratoire de Physique
  Th\'eorique (IRSAMC); F-31062 Toulouse, France
}
\affiliation{
CNRS; LPT (IRSAMC); F-31062 Toulouse, France
}

\date{7 September 2009} 

\begin{abstract} 
We demonstrate how the separation of the total energy of a
self-bound system into a functional of the internal one-body Fermionic
density and a function of an arbitrary wave vector describing the 
center-of-mass kinetic energy can be used to set-up an ``internal'' 
Kohn-Sham scheme.
\end{abstract}


\maketitle 

%
%
\section{Introduction}

Density Functional Theory (DFT) \cite{Dre90,Koh99,DFTLN} is a widely-used 
framework in condensed-matter physics and quantum chemistry to calculate 
properties of many-electron systems, based on the simple local density
instead of the less tractable $N$-body wave function. 

One of the pillars of DFT is the Hohenberg-Kohn (HK) theorem \cite{Hoh64}, 
which in its original form proves that for any non-degenerate system 
of Fermions or Bosons \cite{Dre90}
put into a local external potential $v_{ext}(\mathbf{r})$, there exists 
a unique functional of the local one-body density $\rho(\vec{r})$ that 
gives the exact ground-state energy when $\rho(\vec{r})$ corresponds to 
the exact ground-state density. A thorough mathematical analysis 
of the foundations of the HK theorem was given by Lieb \cite{Lie83}.
A crucial point is that as the theorem is based on the Ritz variational 
principle, it is valid only for systems described by a normalisable 
wave function \cite{Bran}, 
i.e.\ for which a bound (ground) state exists.
\MBcomm{Various extensions of the HK theorem have been proven, for example 
to spin-density energy functionals, for non-local external potentials, 
for relativistic, for time-dependent, or for superconducting systems 
\cite{DFTLN}.
}
The Kohn-Sham (KS) \cite{Koh65} scheme furthermore provides a straightforward 
method to compute self-consistently the ground-state density in a quantum 
framework, defining the local single-particle potential (i.e.\ the 
non-interacting system) which reproduces the exact ground-state density 
through an auxiliary product state.

\MBcomm{The self-consistent mean-field (SCMF) approaches using
effective interaction that are widely 
used to describe the low-energy structure of atomic nuclei \cite{Ben03},
resemble a KS scheme in many ways. Originally conceived as 
Hartree-Fock (HF) of Hartree-Fock-Bogoliubov (HFB) 
method based on an effective in-medium interaction, this framework has 
often been characterized as "nuclear DFT"
\cite{Bei74a,Pet89a,Fay00a,Lal03a,Yu03a,Fur05a,Vre05a}. The similarities
become particularly obvious when the effective interaction is explicitely 
constructed as an energy functional depending on various local densities 
and currents \cite{Car08a}.
There are, however, important conceptual differences that 
prevent the straightforward mapping of the existing nuclear SCMF schemes
onto the standard KS formalism for electronic systems. For example, nuclei 
are self-bound, the intrinsic nuclear states obtained by SCMF methods 
often break several symmetries of the nuclear Hamiltonian, and many extensions of the 
nuclear SCMF method aim at the explicit calculation of correlation effects
instead of absorbing them into the functional.
The present article addresses the first of these points, aiming at a 
KS scheme for self-bound systems. Similar efforts leading to
approximate KS schemes have been made before \cite{Eng07,Bar07}.
Here, we propose an alternative demonstration of the HK theorem
that carefully considers the technical issues arising from 
the separation of internal and center-of-mass coordinates required
for for its application to self-bound systems, and that leads to 
an internal KS scheme.\footnote{\MBcomm{We follow here the nomenclature 
of \cite{Gir08b} to refer to coordinates independent of the center-of-mass 
as "internal" ones, whereas we reserve "intrinsic" for symmetry-breaking 
states out of which bands of rotational states and/or parity vibrations 
can be modeled.} }
}

%
%
\section{The Problem}

\subsection{The role of the external potential}

In electronic systems, the wave function and density are defined in 
the frame attached to the center-of-mass (c.m.) of the 
%
%
\old{nuclei, the latter providing the natural
external potential $v_{ext}(\mathbf{r})$.
The repulsive nature of the electron-electron interaction 
makes the presence of $v_{ext}(\mathbf{r})$ compulsory
in order to reach bound states.
}
\MBcomm{atomic} 
nuclei. The latter also provide naturally the external potential 
$v_{ext}(\mathbf{r})$, whose presence is compulsory to bind
electrons that repel each other.
%
%
The key point of the HK theorem  \old{, though,} 
is that the pure electronic problem 
is universal, whatever the external field (provided it gives a bound state).

In self-bound systems (as atomic nuclei or He droplets)
the situation is intrinsically different because
the net Fermion-Fermion (or Boson-Boson) interaction
%
%
is attractive.
Thus, external fields are not necessary to obtain bound states, 
such that we are immediately in the corresponding "pure" system, with 
the big difference 
%
%
that such systems physically exist.
%
%
\old{
As a consequence, isolated self-bound systems
are plagued by a center-of-mass (c.m.) problem.
}
\MBcomm{
The absence of an external potential, however, has as a consequence
that the modeling of isolated self-bound systems is plagued by 
a center-of-mass (c.m.) problem.
}
%
%
For any stationary state with arbitrary total 
momentum $\vec{P}$, the c.m.\ will be delocalized 
\MBcomm{and evenly distributed}
in the whole space.
\old{
thus the laboratory wave function is not normalizable. 
}
\MBcomm{Even more critical is that such laboratory wave function 
is not normalizable.
}
This prevents
any attempt to formulate DFT for isolated self-bound 
systems in terms of the laboratory density by simply taking 
the limit $v_{ext}(\mathbf{r}) \to 0$ in the HK theorem.
Indeed, this density is an indeterminate constant \cite{Kre01,Eng07},
which forbids to construct a universal functional from it. 
It is of course the ``internal density'' $\rho_{int}$, i.e.\ the 
density relative to the system's c.m., which is of interest.
But standard DFT concepts as formulated so far are not applicable 
yet in terms of a well-defined internal density.

%
%
\subsection{The center-of-mass problem}

A second key point is that in a Hamiltonian and wave-function based 
description of an isolated self-bound system, the Hamiltonian should 
be explicitely translationally invariant to ensure Galilean invariance 
of the wave function
\footnote{
\MBcomm{
Translational invariance, which states that the observables do not depend 
on the position of the c.m., is a necessary, but not sufficient condition 
for the more fundamental Galilean invariance, which ensures that observables
are the same in all inertial frames. In case of a relativistic description
of the quantum $N$-body system \cite{Vre05a}, Lorentz invariance has to be 
considered instead of Galilean invariance.
}
}.
Hence, the $N$-body wave function $\psi$ can be separated into a wave function 
$\Gamma$ that depends on the position 
$\vec{R}=\frac{1}{N}\sum_{j=1}^{N}\vec{r}_j$ of 
the c.m. only, and an "internal" wave function 
$\psi_{int}$ that depends on the remaining ($N-1$) Jacobi 
coordinates ${\boldmath{\xi}}_\alpha$
defined as $\mathbf{\xi}_{1} = \mathbf{r}_2-\vec{r}_1$, 
$\mathbf{\xi}_2=\mathbf{r}_3-\frac{\vec{r}_2+\vec{r}_1}{2}$, \ldots,
$\mathbf{\xi}_{N-1} = \frac{N}{N-1} \, (\vec{r}_N - \vec{R})$
\begin{equation}
\label{eq:psi}
\psi(\vec{r}_1, \ldots , \vec{r}_N) 
= \Gamma(\vec{R}) \; 
  \psi_{int} ({\boldmath{\xi}}_1, \ldots , {\boldmath{\xi}}_{N-1})
\; .
\end{equation}
The $\Gamma (\vec{R})$ describes the motion of the isolated system as 
a whole in any chosen 
inertial frame of reference (as the laboratory).
The $\psi_{int}$ describes the internal properties and is function of the ($N-1$) Jacobi 
coordinates. Of course it could also be written as a function of the $N$ coordinates $\mathbf{r}_i$,
but one of them would be redundant \MBcomm{\cite{Die96}.}


In this context $\rho_{int}$, rather than the laboratory density $\rho$, 
thus becomes the natural quantity on which to construct DFT in a self-bound 
system. It is to be noted that for such a finite system 
it is impossible to construct a
product state that has 
the required structure of $\psi_{int}$. 
In a \old{Hartree-Fock (HF)} \MBcomm{HF} framework, one approximates instead
directly $\psi(\vec{r}_1, \ldots, \vec{r}_N)$ by a Slater determinant
in $N$ coordinates $\vec{r}_i$ in the c.m.\ frame of the system.
Consequently, the HF state contains (at least) \old{three} one
redundant coordinate, that introduces a spurious coupling between the
internal \old{state} properties and the c.m. motion \cite{RS80}.
For this reason, the HF approximation 
sacrifices ``Galilean invariance for the sake of the Pauli principle'', 
to quote \cite{Schm01a}. A rigorous remedy is to perform projected HF, 
where projection before variation on c.m.\ momentum restores 
Galilean invariance at the price of abandoning the independent-particle 
description \cite{Schm01a,Rod04a}.
This reasoning does not hold, in principle,
for DFT, where the key ingredient 
is the density, not an explicit $N$-body wave function.
%

\JMcomm{A} demonstration of a rigorous internal HK theorem
has been made recently
\JMcomm{in \cite{Eng07,Bar07} (by two different ways),}
aiming at the correct 
separation of the internal properties from the c.m.\ motion, 
but none of them led to a rigorous internal KS scheme.
A source term coupled to the $N$-body internal density operator was introduced 
in \cite{Eng07}, allowing to express the exact total energy of a 
self-bound system as a functional of this operator.
A scheme to construct a corresponding non-interating system 
in a systematic manner was proposed, but its link with the KS scheme of
traditional DFT remains unclear.
%
%
\JMcomm{
In \cite{Bar07}, it was shown that the internal energy of a self-bound 
system can be written as a functional of the internal one-body density
and an approximate KS scheme was proposed, valid only if the 
c.m.\ coordinate is treated as an adiabatic variable.
%
%
A different approach to the problem is taken in 
Refs.~\cite{Gir08b,Gir08c}, where a (non-translationally invariant) oscillator potential that traps the 
center-of-mass is added to the self-bound Hamiltonian.
This aproach has the particular characteristic that 
\old{it leaves the internal properties unchanged. ( }
\MBcomm{it does not affect the internal properties of the system:}
the ground-state wave function is a wave packet that \old{still}
factorizes into the form of Eq.~(\ref{eq:psi}), $\Gamma(\vec{R})$ now being \old{normalizable - a Gaussian).}
\MBcomm{a Gaussian and with that normalizable.}
The laboratory density $\rho$ is then well defined 
and a KS scheme for $\rho$ can be rigorously set up.
The internal density $\rho_{int}$ can be deduced from $\rho$ by deconvolution.
\old{
But the obtained energy functional / KS equations does not consist in an internal energy functional / KS scheme,
thus is not directly comparable to self-consistent 
mean-field like calculations that use effective interactions.}
\MBcomm{
However, the thus obtained energy functional and KS equations are neither 
an internal energy functional nor internal KS equations.
}
Thus, the question of a rigorous formulation of an internal KS scheme
\MBcomm{comparable to SCMF calculations using an effective interaction}
remains open. Here, we propose a complementary way than those found
in \cite{Eng07,Bar07} to demonstrate the
internal HK theorem, whose link with the traditional HK theorem is more clear.
This directly leads to the formulation of a general internal KS scheme.
}
%
%
\section{DFT in internal degrees of freedom}
%
%
\subsection{Separation of internal and c.m.\ coordinates}

We start from a general translationally invariant $N$-body Hamiltonian
composed of the usual kinetic energy term, and a 
%
%
two-body potential $u$ which describes the Fermion-Fermion (or Boson-Boson) 
interaction 
\begin{equation}
\label{eq:H}
H 
=   \sum_{i=1}^{N} \frac{\vec{p}^2_i}{2m} 
  + \sum_{\stackrel{i,j=1}{i > j}}^{N} u (\vec{r}_i-\vec{r}_j) 
\; .
\end{equation}
\MBcomm{For the sake of simplicity of the demonstration we assume a 
momentum-independent 2-body interaction and $N$ identical particules.
The generalization to 3-body interactions is straightforward,
the generalization to systems containing 
different types of particles will be discused elsewehere.}
%
We rewrite the Hamiltonian using the $N-1$ Jacobi coordinates
$\xi_\alpha$, to decouple the internal properties from the c.m.\ motion. 
The $\xi_\alpha$ are 
\old{independent from $\vec{R}$ and relative to the c.m.\ of the other 
$1, \ldots, i-1$ Fermions, and
} 
to be distinguished from the $N$ 
"laboratory coordinates" $\vec{r}_i$,
and the $N$ "c.m.\ frame
coordinates" $(\vec{r}_i-\vec{R})$ relative to the total c.m. $\vec{R}$.
One can then separate (\ref{eq:H}) into $H = H_{CM} + H_{int}$, where
$H_{CM} = -(\hbar^2/2M) \Delta_\vec{R}$ (with $M = Nm$ being the total mass)
is a one-body operator acting in $\vec{R}$ space only, and
$H_{int}$ is a $(N-1)$ body 
operator in the $\xi_\alpha$ space. $H_{int}$ contains the 
%
%
interaction $u$ which can be rewritten as a function of the 
$\xi_\alpha$ [which we denote $u(\{\xi_\alpha\})$ for simplicity],
and the internal kinetic energy, which is expressed in
terms of the conjugate momentum $\tau_\alpha$ of $\xi_\alpha$ and the corresponding reduced mass
$\mu_\alpha = m\frac{\alpha}{\alpha+1}$.
As $[H_{CM},H_{int}]=0$, the eigenstate $\psi$ of $H$ can be build
\old{constructed} as a \old{direct} product of the form (\ref{eq:psi}) with
\begin{eqnarray}
\label{eq:hcm}
-\frac{\hbar^2}{2M} \Delta_\vec{R} \Gamma 
& = & E_{CM} \Gamma \, ,
\\
\label{eq:hint}
H_{int} \psi_{int} 
& = & E_{int} \psi_{int} \, , 
\end{eqnarray}
where
\begin{equation}
H_{int}
= \sum_{\alpha=1}^{N-1} \frac{\tau_\alpha^2}{2\mu_\alpha} + u(\{\xi_\alpha\})
\; .
\end{equation}
There is no bound state for $\Gamma(\vec{R})$ as the solutions are arbitrary 
stationary plane waves, leading to arbitrary c.m.\ energy 
\old{$E_{CM}=\frac{\hbar^2\mathbf{K}^2}{2M}$}
\MBcomm{$E_{CM} = \hbar^2\mathbf{K}^2/(2M)$}
and delocalization of $\vec{R}$. We will come back to the interpretation
of $\Gamma(\vec{R})$ below. By definition of a self-bound system,
$\psi_{int}$ is a bound, thus normalizable, state.
The correponding total energy $E = E_{CM}+E_{int}$ splits into
\begin{eqnarray}
\label{eq:E0}
E^{(\mathbf{K})}[\psi_{int}] 
& = & \frac{\hbar^2\mathbf{K}^2}{2M}
      + \frac{E_{int} [\psi_{int}]}{(\psi_{int}|\psi_{int})} \\
\label{eq:E00}
E_{int}[\psi_{int}]
& = & (\psi_{int}| H_{int} |\psi_{int})
\end{eqnarray}
where the internal energy
is obviously a functional of $\psi_{int}$,
and the c.m. energy is parametrized by an arbitrary $\mathbf{K}$.
We see that the c.m.\ properties (given by $\mathbf{K}$) and the internal properties
(given by $\psi_{int}$) are fully decoupled.
The ground state $\psi_{int}$ of $H_{int}$ is obtained by minimization of the total energy $E^{(\mathbf{K})}[\psi_{int}]$
for a given $\mathbf{K}$, or equivalently of $E_{int}[\psi_{int}]$ imposing normalisation.

The previous steps have allowed to uniquely identify and separate 
the c.m.\ motion.
\old{, which is a  key issue for self-bound systems.} 
In traditional electronic DFT the problem does not show up as the
electronic properties are defined in the frame attached to the c.m. 
of the nuclei, where the nuclear background is accounted for by introducing an 
external local one-body potential $v_{ext}(\vec{r})$ that provides the key 
ingredient of the HK theorem. 
%
%
In the case of a self-bound system, $v_{ext}$ is not compulsory.
To faciliate the proof of the HK theorem, however, we introduce an arbitrary
potential $v_{aux}$, which serves as a mathematical auxiliary
and can be safely dropped at the end to recover an isolated self-bound system.

%
%
\subsection{A translational invariant auxiliary potential}

\old{But mind that if we want to preserve c.m.\ properties}
\MBcomm{To conserve the separation of c.m.\ and internal properties,} 
we cannot simply use a  one-body potential of the form 
$v_{aux}({\bf r})$. The potential $v_{aux}$ should necessarily 
verify two conditions:
(1) translational invariance and
(2) as we are interested in the internal properties, the redundant 
c.m.\ coordinate should be removed (as discussed previously).
These two conditions impose the form $\sum_{i=1}^{N} 
v_{aux}(\mathbf{r}_i-\mathbf{R})$ as already used in 
\cite{Eng07,Bar07},
which corresponds to an arbitrary potential seen in the c.m. frame.
\old{$N$-body potential in the c.m.\ frame} 
It can  be expressed as a function of the 
Jacobi coordinates only, $\sum_{i=1}^{N} 
v_{aux}(\mathbf{r}_i-\mathbf{R}) = v_{aux}(\{\xi_\alpha\})$;
hence, it does not couple to c.m.\ properties, 
the decomposition (\ref{eq:psi}) for $\psi$ still holds 
with ${H_{int}} \rightarrow H_{int} + v_{aux}(\{\xi_\alpha\})$ in (\ref{eq:hint}).
Of course, the associated internal wave function is 
modified accordingly,
and consequently all internal observables,
but for the sake of simplicity we keep the same 
notations ($\psi_{int}, E_{int},H_{int}$).

\old{To} 
\MBcomm{As the next step, we}
evaluate the contribution of the auxiliary potential term 
$(\psi_{int}|v_{aux}(\{\xi_\alpha\})|\psi_{int})$ to the internal 
energy. 
\MBcomm{First, we}
note that for any \JMcomm{operator $\hat{f}$}
\MBcomm{that can be expressed}
through the Jacobi coordinates in position representation
\JMcomm{(we note $\hat{f}(\{\xi_\alpha\})$ when expressed through the Jacobi coordinates
and $\hat{f}(\{\vec{r}_i\})$ when expressed through the laboratory coordinates),}
we have the relation
\begin{eqnarray}
\label{eq:rel}
\lefteqn{(\psi_{int}| \hat{f}(\{\xi_\alpha\}) |\psi_{int})}\nonumber
\\
& = & \int \! d\mathbf{\xi}_1 \cdots  d\mathbf{\xi}_{N-1} \, \psi_{int}^* (\{\xi_\alpha\}) \hat{f}(\{\xi_\alpha\}) \, 
\psi_{int} (\{\xi_\alpha\})
\nonumber\\
& = & \int \! d\mathbf{R} \; d\mathbf{\xi}_1 \cdots  d\mathbf{\xi}_{N-1} \, 
      \delta(\mathbf{R}) \, \psi_{int}^* (\{\xi_\alpha\}) \hat{f}(\{\xi_\alpha\}) \, 
      \psi_{int} (\{\xi_\alpha\})
\nonumber\\
& = & \int \! d\vec{r}_1 \cdots d\vec{r}_{N} \, 
      \delta(\mathbf{R}) \, 
      \psi_{int}^* (\{\vec{r}_i\}) \hat{f}(\{\vec{r}_i\}) \, \psi_{int}(\{\vec{r}_i\})
\, .
\end{eqnarray}
%
%
We see that the "internal mean values" calculated with $\psi_{int}$
expressed as a function of the $(N-1)$ $\{\xi_\alpha\}$
can also be calculated with $\psi_{int}$
expressed as a function of the $N$ coordinates $\{\mathbf{r}_i\}$.
The transformation from the $\{\xi_\alpha\}$ to the $\{\mathbf{r}_i\}$ introduces 
a $\delta(\mathbf{R})$ that represents the dependence of the 
redundant coordinate on the others 
\footnote{
More generally, we can introduce a $\delta(\mathbf{R}-\mathbf{a})$ where
$\mathbf{a}$ is an arbitrary translation vector, which
\JMcomm{represents the position of the system's c.m.\  in .the laboratory 
coordinates $\{ \vec{r}_i \}$.
For sake of simplicity of the notation, we chose $\mathbf{a}=\mathbf{0}$ without loss of generality.}
}.

\MBcomm{For the mean value of the auxiliary potential,}
relation (\ref{eq:rel}) leads to
\begin{eqnarray}
\label{eq:Eext}
\lefteqn{
(\psi_{int}|v_{aux}(\{\xi_\alpha\})|\psi_{int}) 
} \nonumber\\
& = & \int \! d\vec{r}_1 \cdots d\vec{r}_N \; 
      \delta(\mathbf{R}) |\psi_{int}(\{\vec{r}_i\})|^2 \, 
      \sum_{i=1}^N v_{aux}(\vec{r}_i - \vec{R}) 
      \nonumber\\
& = & \sum_{i=1}^N 
      \int \! d\eta \; v_{aux}(\eta) 
      \int \! d\vec{r}_1 \cdots d\vec{r}_N \; 
      |\psi(\vec{r}_1, \ldots, \vec{r}_{N})|^2  
      \nonumber\\
&   & \quad \times \delta \big( \eta-(\vec{r}_i-\vec{R}) \big) 
      \nonumber\\
& = & \sum_{i=1}^N 
      \int \! d\vec{r} \; v_{aux}(\vec{r}) \, 
      \, \frac{\rho_{int}(\vec{r})}{N} 
      \nonumber\\
& = & \int \! d\vec{r} \; v_{aux}(\vec{r}) \, \rho_{int}(\vec{r})
\,
\end{eqnarray}
where we have introduced the \textit{internal} density
\begin{eqnarray}
\label{eq:rho_int0}
\lefteqn{\rho_{int}(\vec{r})/N} \\
& = & \int \! d\vec{r}_1 \cdots d\vec{r}_N \; 
      \delta(\mathbf{R}) |\psi_{int}(\{\vec{r}_i\})|^2 \, 
      \delta \big( \vec{r} - (\vec{r}_i-\vec{R}) \big)
      \nonumber \\
& = & \int \! d\vec{r}_1 \cdots d\vec{r}_N \; 
      \delta(\mathbf{R}) |\psi_{int}(\{\vec{r}_i\})|^2 \, 
      \delta \big( \vec{r} - (\vec{r}_N-\vec{R}) \big)
      \nonumber \\
& = & \Big(\frac{N}{N-1}\Big)^3 
      \int \! d\mathbf{R} d\vec{\xi}_1 \cdots  d\mathbf{\xi}_{N-1} \; 
      \delta(\mathbf{R})
      \nonumber \\
&&    \hspace{1.7cm} \times \big| \psi_{int} \big(\{\xi_\alpha\}\big) \big|^2 \delta(\mathbf{\xi}_{N-1}-\tfrac{N\vec{r}}{N-1})
      \nonumber \\
& = & \Big(\frac{N}{N-1}\Big)^3 
      \int \! d\vec{\xi}_1 \cdots  d\mathbf{\xi}_{N-2} \; 
      \big| \psi_{int} \big(\mathbf{\xi}_1, \ldots, \mathbf{\xi}_{N-2},
                   \tfrac{N\vec{r}}{N-1} \big) \big|^2
      \nonumber .
\end{eqnarray}
The density $\rho_{int}(\vec{r})$ is normalized to $N$ and a 
function of the c.m.\ frame coordinates. The 
laboratory density is obtained by convolution with the c.m.\ wave 
function as in \cite{Gir08b,Kaz86}. The potential $v_{aux}$ 
%
%
that is $N$ body 
with respect to the laboratory coordinates (and $(N-1)$ body when 
expressed with Jacobi coordinates), becomes one body (and local) when
expressed with the c.m. frame coordinates. 
The energy 
$E_{int}[\psi_{int}]$ (\ref{eq:E00}), and thus the total energy $E^{(\mathbf{K})}[\psi_{int}]$ (\ref{eq:E0}),
are then to be complemented by
\begin{equation}
\label{eq:newE}
E_{int}
\rightarrow E_{int} + \int \! d\vec{r} \, v_{aux}(\vec{r}) \rho_{int}(\vec{r})
\, .
\end{equation}
%

%
%
\subsection{The Hohenberg-Kohn theorem}

\MBcomm{The internal energy}
${E}_{int}$ \old{of the new total energy, Eq.~(\ref{eq:newE}),}
remains obviously a functional of ${\psi}_{int}$. As in its definition enters 
an arbitrary one-body potential in the c.m. frame of the form 
$\int \! d\vec{r} \; v_{aux}(\vec{r}) \, \rho_{int}(\vec{r})$,
and as the ground state of $H_{int}$ is obtained by minimization of ${E}_{int}$,
we can directly apply 
the usual proof of the HK theorem \cite{Hoh64}
and claim that for a 
non-degenerate 
ground state ${\psi}_{int}$,
and for a given Fermion or Boson type (i.e.\ a given interaction $u$),
the internal energy ${E}_{int}$ of a self-bound 
system, Eq.~(\ref{eq:newE}), can be expressed as a unique
functional of  ${\rho}_{int}$. 
%
%
As already emphasized, the HK theorem is valid only for arbitrary ``external'' potentials 
that lead to \textit{bound} ground states \cite{Lie83}. As a direct consequence, 
the internal DFT scheme is valid only for potentials 
$v_{aux}$ that lead to \textit{bound} internal ground states $\psi_{int}$. 
%
%
As for self-bound systems, described by our formalism at the limit $v_{aux}\rightarrow 0$,
$\psi_{int}$ should by definition be a bound ground state,
the previous conclusions still hold
without breaking the consistency of the scheme.
\old{(because the fermion-fermion interaction $u$ should be sufficiently attractive)}
\old{for many combinations of neutron and proton numbers}
\old{i.e. the electronic case,} 
\old{This is a major difference with the \old{traditional} electronic DFT case,
where the electron-electron repulsion prevents us to preserve the DFT conclusions in the limit $v_{ext}\rightarrow 0$.}
\old{because no bound state then can  occur.} 
\old{The form (\ref{eq:Eint}) for the internal energy allows clearly to apply 
directy standard pseudo-potential  tools which would `absorb' 
the one-body `core' nucleons wavefunction on a pseudo-potential,
to treat explicitely only the `valence' ones. If the pseudo-potential 
is local, it can be inserted in $v_{int}$.}

%
%
\subsection{The internal Kohn-Sham scheme}

To recover the \old{associated} ``internal'' KS scheme, we assume, as in the
traditional KS scheme, that there exists, \textit{in the c.m.\ frame}, a local 
single-particle potential (i.e. a $N$-body non-interacting system)
which reproduces the density $\rho_{int}$ of the interacting system.
We develop $\rho_{int}$ in the corresponding
\old{unique}
basis $\varphi^i_{int}$ of one-body orbitals expressed in c.m.\ frame 
coordinates $\vec{r}$
\footnote{
\JMcomm{
Even if only ($N-1$) coordinates are sufficient to describe the internal properties,
we still deal with a system of $N$ particles.
Thus, we have to introduce $N$ orbitals in the KS scheme
if we want them to be interpreted (to first order) \old{only)} as single-particle orbitals
and obtain a scheme comparable to \old{mean-field like} \MBcomm{SCMF} calculations 
\old{with} \MBcomm{using} effective interactions.
}
}
\begin{eqnarray}
\rho_{int}(\vec{r}) 
= \sum_{i=1}^N \big| \varphi^i_{int}(\vec{r}) \big|^2
\, .
\end{eqnarray}
The KS assumption implies $\varphi^i_{int}[\rho_{int}]$ \cite{Dre90}; hence,
we can rewrite \old{the total energy seen in the internal frame} $E_{int}$ as
\footnote{
\MBcomm{To keep close contact with standard DFT, we make here the usual
separation of the energy into direct (Hartree) and exchange-correlation
parts. Owing to the complexity of the nucleon-nucleon interaction in the 
vacuum, strong correlations in the nuclear medium, and the appearance of 
three-body forces, it common practice in nuclear applications to construct
approximate expressions for the entire functional. This, however, does not
affect the conclusions of the present paper.
}
}
\old{Eq.~(\ref{eq:E1}), as} 
%
%
\old{(we keep $v_{aux}$ for generality, but recall that 
it can be put to zero in the self-bound case)}
\begin{eqnarray}
\label{eq:Eint}
E_{int}[\rho_{int}] 
& = & \sum_{i=1}^{N} (\varphi^i_{int}|\frac{\vec{p}^2}{2m}|\varphi^i_{int}) 
      + E_{H}[\rho_{int}] \\
&   & + E_{XC}[\rho_{int}] 
      + \int \! d\vec{r} \; v_{aux}(\vec{r}) \, \rho_{int}(\vec{r}) 
      \nonumber\\
E_{XC}[\rho_{int}] 
& = & \frac{1}{2} \int \! d\vec{r} \, d\vec{r'} \, 
      \gamma_{int}(\vec{r},\vec{r'}) \, 
      u(\vec{r}-\vec{r'}) - E_{H}[\rho_{int}]  
      \nonumber\\
&   & + (\psi_{int}|\sum_{\alpha=1}^{N-1} \frac{\tau_\alpha^2}{2\mu_\alpha}|\psi_{int}) 
      - \sum_{i=1}^{N} (\varphi^i_{int}|\frac{\vec{p}^2}{2m}|\varphi^i_{int}) 
\nonumber
\end{eqnarray}
where we added and subtracted the internal Hartree energy
$E_{H}[\rho_{int}] = \frac{1}{2} \int \! d\vec{r} \, d\vec{r'} \, 
\rho_{int}(\vec{r}) \, \rho_{int}(\vec{r'}) \, u(\vec{r}-\vec{r'})$
for the direct part and the non-interacting kinetic energy
$\sum_{i=1}^{N} (\varphi^i_{int}|\frac{\vec{p}^2}{2m}|\varphi^i_{int})$.
For convenience, we introduced the local part of the two-body 
\textit{internal} density matrix
\begin{eqnarray}
\label{eq:gamint0}
\lefteqn{\gamma_{int}(\vec{r},\vec{r'})}
      \\
& = & \int \! d\vec{r}_1 \cdots d\vec{r}_N \; 
      \delta(\mathbf{R}) |\psi_{int}(\{\vec{r}_i\})|^2 \,
      \nonumber \\
&   & \hspace{1.cm} \times 
      \delta \big( \vec{r} - (\vec{r}_i-\mathbf{R}) \big)
      \delta \big( \vec{r'} - (\vec{r}_{j\ne i}-\mathbf{R}) \big)
      \nonumber \\
& = & \frac{N(N-1)}{2} \Big(\frac{N-1}{N-2}\Big)^3 \Big(\frac{N}{N-1}\Big)^3
      \int \! d\mathbf{\xi}_1 \cdots  d\mathbf{\xi}_{N-3}
      \nonumber\\
&   & \hspace{1.cm} \times \Big| \psi_{int} \Big(\mathbf{\xi}_1, \ldots, \mathbf{\xi}_{N-3}, 
      \tfrac{\vec{r'}+(N-1)\vec{r} }{N-2},\tfrac{N\vec{r'}}{N-1} \Big) \Big|^2
      \nonumber 
\end{eqnarray}
(using similar steps as in (\ref{eq:rho_int0}) and
$\vec{r}_{N}-\mathbf{R}=\frac{N-1}{N}{\xi}_{N-1}$
and
$\vec{r}_{N-1}-\mathbf{R}=\frac{N-2}{N-1}{\xi}_{N-2}-\frac{{\xi}_{N-1}}{N}$
%
%
).
$\gamma_{int}$ has the required normalisation to $N(N-1)/2$,
is a function of the c.m.\ frame coordinates and gives 
\MBcomm{the two-body density matrix}
$\gamma(\vec{r},\vec{r'})$ 
\MBcomm{in the laboratory} by convolution with the c.m.\ wave 
function $\Gamma (\vec{R})$. Applying Eq.~(\ref{eq:rel}) to the 
$u(\{\xi_\alpha\})$ part of $H_{int}$ and inserting Eq.~(\ref{eq:gamint0})
gives directly 
\begin{equation}
(\psi_{int}| u(\{\xi_\alpha\}) |\psi_{int})
= \frac{1}{2} \int \! d\vec{r} \, d\vec{r'} \; 
  \gamma_{int}(\vec{r},\vec{r'}) \; u(\vec{r}-\vec{r'})
\; .
\end{equation}
Following similar steps as in Eq.~(\ref{eq:rel}), one can show that the
interacting kinetic energy can be rewritten as
\begin{eqnarray}
\label{eq:intkin}
\lefteqn{
(\psi_{int}| 
\sum_{\alpha=1}^{N-1} \frac{\tau_\alpha^2}{2\mu_\alpha}|\psi_{int})
} \\
& = & (\psi_{int}| -\frac{\hbar^2\Delta_\vec{R}}{2M} +
\sum_{\alpha=1}^{N-1} \frac{\tau_\alpha^2}{2\mu_\alpha}|\psi_{int})
\nonumber \\
& = & \int \! d\vec{r}_1 \cdots d\vec{r}_N \delta(\mathbf{R}) \,
  \psi_{int}^*(\{\vec{r}_i\})
  \sum_{i=1}^N \frac{\mathbf{p}_i^2}{2m}\psi_{int}(\{\vec{r}_i\})
\nonumber 
\, .
\end{eqnarray}
%
\MBcomm{Equation~(\ref{eq:intkin})}
makes it clear that the difference
\MBcomm{to}
the non-interacting kinetic energy 
$\sum_{i=1}^{N} \int d\vec{r} \varphi^{i*}_{int}(\vec{r})
\frac{\vec{p}^2}{2m}\varphi^i_{int}(\vec{r})$
comes, on the one hand, from the correlations neglected in the 
traditional independent-particle framework, but also from the c.m.\ 
correlations (the $\delta(\mathbf{R})$ term in the previous expression).
The inclusion of the c.m.\ correlations in the functional is the main 
difference 
\old{with}
to the traditional KS scheme.
\old{A}
Still, the key point is that the internal pure
exchange-correlation energy $E_{XC}[\rho_{int}]$ can be expressed as a
functional of $\rho_{int}$.
\old{Even if in practice one will have to work with an approximate 
functional for $E_{XC}[\rho_{int}]$, this will not break Galilean
invariance because we separated properly the internal motion from that
of the c.m. }
Varying $E_\mathbf{K}[\rho_{int}]$, Eq.~(\ref{eq:E0}), or 
equivalently $E_{int}[\rho_{int}]$, 
Eq.~(\ref{eq:Eint}), with respect to $\varphi^{i*}_{int}$, and imposing 
normality of the $\{\varphi^i_{int}\}$,
\begin{eqnarray}
\frac{\delta}{\delta \varphi^{i*}_{int} (\vec{r})}
\Big(   E_{int}[\rho_{int}] 
      - \sum_{i=1}^N \epsilon_i \, (\varphi^i_{int}|\varphi^i_{int}) 
\Big) 
& = & 0
\, ,
\end{eqnarray}
leads to "internal" Kohn-Sham equations for the $\{\varphi^i_{int}\}$
\begin{equation}
\label{eq:varphi_i}
\Big(
- \frac{\hbar^2}{2m}\Delta 
+ U_H[\rho_{int}] 
+ U_{XC}[\rho_{int}] 
+ v_{aux}
\Big) \varphi^i_{int} = \epsilon_i \varphi^i_{int}
\end{equation}
with $U_{H}[\rho_{int}](\vec{r}) 
= \delta E_{H}[\rho_{int}] / \delta \rho_{int}(\vec{r})$ and 
$U_{XC}[\rho_{int}](\vec{r}) 
= \delta E_{XC}[\rho_{int}] / \delta \rho_{int}(\vec{r})$, which is 
local as expected.
Equation~(\ref{eq:varphi_i}) has the same form as the traditional KS 
equations formulated \old{in the laboratory frame} for non-translationally 
invariant Hamiltonians \cite{Koh65}.
Here, however, we have justified its \old{form} use in the c.m.\ frame for 
self-bound systems described with translational-invariant Hamiltonians,
and shown that the functional form of $U_{XC}[\rho_{int}] $ differs by the 
inclusion of c.m.\ correlations. Together with Eqs.~(\ref{eq:E0}) 
and (\ref{eq:Eint}), Eq.~(\ref{eq:varphi_i}) defines
completely the total energy $E_\mathbf{K}[\rho_{int}]$ as the sum of
the c.m.\ kinetic energy and of the internal energy.

%
%
\subsection{The laboratory density}

\old{Finally, it} 
It is instructive to calculate the laboratory density $\rho$. Following
Ref.~\cite{Gir08b}, one obtains
\begin{equation}
\rho(\vec{r}) = \int \! d\vec{\mathbf{R}} \, |\Gamma(\vec{R})|^2 \, 
\rho_{int}(\vec{r}-\vec{R})
\, .
\end{equation} 
As $\Gamma (\vec{R})$ is a plane wave, 
\old{$|\Gamma(\vec{R})|^2$ is a constant; hence,}
$\rho(\vec{r})$ is constant. This confirms that the 
usual definition of the "laboratory density" lacks a meaningful 
interpretation for isolated self-bound systems. Of course this full 
delocalisation does not occur in an experiment because observed self-bound 
systems are not \textit{isolated} anymore. 
The observables related to $\Gamma (\vec{R})$, 
i.e.\ position, momentum or kinetic energy, 
are used to transform all observables into the 
c.m. frame, thereby explicitely using the Galilean invariance.
The key point is that the decoupling of the c.m.\ motion allows to deduce 
internal properties preserving Galilei invariance, whereas 
$\Gamma (\vec{R})$ is left to the choice of experimental conditions.
\old{The limit of the usual DFT is recovered fixing the c.m.\ in the
laboratory frame, i.e.\ constraining $|\Gamma(\mathbf{R})|^2 = \delta(\mathbf{R})$, which
\old{makes obsolete Eq.~(\ref{eq:gamma}) and} 
gives $\rho(\vec{r})=\rho_{int}(\vec{r})$ \JMcomm{and $\mathbf{K}=0$}
(it behaves as $v_{aux}(\vec{r}_i - \vec{R})\rightarrow v_{aux}
(\vec{r}_i - 0)\rightarrow v_{ext} (\vec{r}_i)$ in usual DFT notations).}

%
%
\section{Summary and Conclusions}

In summary, we have shown
\old{\JMcomm{in a complementary way than those found}}
\MBcomm{in a way complementary to those proposed}
\JMcomm{in Refs.~\cite{Eng07,Bar07}}
that the total energy of a self-bound system 
can be expressed as a functional of
\old{the one-body c.m.\ wave function $\Gamma$ and}
the total one-body internal density $\rho_{int}$.
\old{and is parametrized by by the c.m.\ wave vector $\mathbf{K}$.} 
\MBcomm{The energy in the laboratory frame contains the c.m.\ wave 
vector $\mathbf{K}$ as a parameter that can be freely chosen according 
to experimental conditions.} 
\JMcomm{Then, we have shown rigorously that}
the internal properties of the system are described by an internal 
KS scheme.
\JMcomm{
The key difference \old{with} \MBcomm{to} the traditional HK/KS functional 
is the inclusion of \old{the} c.m.\ correlations.
}
The question about the universal validity of the Kohn-Sham hypothesis, known as the 
``non-interacting v-representability'' problem \cite{Dre90}, however, remains to 
be answered, as in traditional DFT.
\old{
A more detailed analysis, particularly of the case 
\old{of interest in} of nuclear physics, where a very similar
phenomenologically motivated framework is widely used \cite{Ben03}, 
is underway
}
\old{
\JMcomm{We should 
stress already here that the functional $E_{int}$ cannot be
the HF expectation value of a force, neither is there a need for a
c.m.\ correction, as commonly used in "nuclear DFT" \cite{Ben03}.}}
\old{Our internal DFT scheme} 
\MBcomm{The internal DFT scheme proposed here}
provides \old{also} a justification for the
application of DFT to isolated $^3$He and $^4$He droplets \cite{Bar06}.
\old{Finally, the same reasoning should be applied
to rotational invariance to formulate the theory in term of the so-called 
"intrinsic" one-body density \cite{Gir08a} (which is not directly observable 
and is more complicated because rotation does not decouple from internal 
motion).}
\MBcomm{The present paper establishes also the first step towards an
Kohn-Sham scheme applicable to nuclear structure physics.
Further necessary developments are the generalization to two (or more) 
species of interacting particles, and the treatment broken rotational
and space-inversion symmetry that requires to formulate the theory in 
term of the so-called "intrinsic" one-body density as defined in 
Ref.~\cite{Gir08a}.
}

%
%
\subsection*{Acknowledgments.}

The authors acknowledge numerous enlightening discussions with J.~Labarsouque, 
J.~Navarro, P.-G.\ Reinhard, and P.~Quentin,
and thank the Institut Universitaire de France and the
Agence Nationale de la Recherche (ANR-06-BLAN-0319-02) for financial support.
One of us (JM) thanks the Centre 
d'Etudes Nucl{\'e}aires de Bordeaux-Gradignan for warm
hospitality.

%
%

\end{document}